\documentclass[prc,preprint,nofootinbib]{revtex4}
\usepackage{amssymb}
\usepackage{longtable}
\usepackage{graphicx}
\usepackage{fleqn}
\usepackage{epsfig}
\usepackage[figuresright]{rotating}

\begin{document}

\preprint{}
\author{P. E. Koehler }
\email{koehlerpe@ornl.gov}
\affiliation{Physics Division, Oak Ridge National Laboratory,Oak Ridge, TN 37831}
\title{Reduced neutron widths in the nuclear data ensemble: Experiment and
theory do not agree}
\date{\today }

\begin{abstract}
I have analyzed reduced neutron widths ($\Gamma _{n}^{0}$) for the subset of
1245 resonances in the nuclear data ensemble (NDE) for which they have been
reported. Random matrix theory (RMT) predicts for the Gaussian orthogonal
ensemble (GOE) that these widths should follow a $\chi ^{2}$ distribution
having one degree of freedom ($\nu =1$) - the Porter Thomas distribution
(PTD). Careful analysis of the $\Gamma _{n}^{0}$ values in the NDE rejects
the validity of the PTD with a statistical significance of at least 99.97\% (%
$\nu =0.801\pm 0.052$). This striking disagreement with the RMT prediction
is most likely due to the inclusion of significant \textit{p}-wave
contamination to the supposedly pure \textit{s}-wave NDE. When an energy
dependent threshold is used to remove the \textit{p}-wave contamination, the
PTD is still rejected with a statistical significance of at least 98.17\% ($%
\nu =1.217\pm 0.092$). Furthermore, examination of the primary references
for the NDE reveals that many resonances in most of the individual data sets
were selected using methods derived from RMT. Therefore, using the full NDE
data set to test RMT predictions seems highly questionable. These results
cast very serious doubt on claims that the NDE represents a striking
confirmation of RMT.
\end{abstract}

\maketitle

\section{Introduction\label{intro}}

The nuclear data ensemble (NDE) \cite{Ha82,Bo83} is a set of 1407 resonance
energies consisting of 30 sequences in 27 different nuclides. The ensemble
was assembled to test predictions of random matrix theory (RMT) \cite{We2009}%
. Fluctuation properties of resonance energies in the NDE were found to be
in remarkably close agreement with RMT predictions for the Gaussian
orthogonal ensemble (GOE). Although there have been several other successful
tests of RMT using nuclear resonances (e.g., \cite{Sh87,Ca94}),the NDE is
perhaps the most important because, as stated in Ref. \cite{We2009}, "As a
result of these analyses, it became generally accepted that proton and
neutron resonances in medium weight and heavy nuclei agree with GOE
predictions." Hence, the NDE routinely is cited as providing striking
confirmation of RMT.

Reduced neutron widths ($\Gamma _{n}^{0}$) have been reported for a subset
of 1245 resonances in the NDE of Ref. \cite{Bo83}, consisting of 14 to 178
measurements for 24 nuclides. If the GOE correctly describes the data, RMT
predicts these widths should follow a $\chi ^{2}$ distribution having one
degree of freedom ($\nu =1$) - the Porter Thomas distribution (PTD). It has
be argued that the PTD is more generally valid \cite{Br81a} and more robust 
\cite{Gr83} than other RMT predictions. Also, it has been demonstrated \cite%
{Co79} that a much more reliable analysis of spectra fluctuations can be
performed using neutron widths than using resonance energies. In addition,
it has been shown \cite{Al92} that resonance width fluctuations are more
sensitive than energy fluctuations to the degree of chaos in model quantum
systems. In addition, it is straightforward \cite{Ko2010} to account for
experimental effects such as missed or spurious resonances, and to use the
statistically efficient maximum-likelihood (ML) method while using width
data, but the same cannot be said for energy data. For these reasons, a test
of the PTD using the NDE neutron widths could be very valuable.

In Ref. \cite{Bo83} such a test on a subset of the NDE data was described
from which it was concluded that there was satisfactory accord between
theory and experiment. However, there are several problems with the analysis
of Ref. \cite{Bo83}, which I will describe below. I find that when the data
are analyzed more carefully, they do not agree with the PTD.

I will show below that the NDE suffers from significant \textit{p}-wave
contamination. For such resonances, reported reduced neutron widths $\Gamma
_{n}^{0}$ are actually only "effective" reduced widths $g\Gamma _{n}/\sqrt{%
E_{n}}$, where $g=\frac{2J+1}{(2I+1)(2j+1)}$ (with $J$, $I$, and $j$ being
spins of the resonance, target, and neutron, respectively) is the spin
statistical factor. All \textit{s}-wave resonances for NDE nuclides included
in my analysis have $J=\frac{1}{2}$ and hence $g=1$ and $g\Gamma
_{n}^{0}=\Gamma _{n}^{0}$. However, there are two resonance spins possible
for \textit{p}-wave resonances for these nuclides; $J$ $=\frac{1}{2}$ or $%
\frac{3}{2}$, and hence $g=1$ or 2. As a reminder of these facts, I will use 
$g\Gamma _{n}^{0}$ when referring to reduced neutrons widths in the
remainder of this paper.

\section{Reconstructing the NDE}

To my knowledge, the resonance energies and neutron widths in the NDE have
never been published as a set. It should be possible, however, to
reconstruct this information from the nuclides and corresponding number of
resonances given in the NDE papers \cite{Ha82,Bo83} and the primary
references cited therein. Unfortunately, several of the references listed in
the NDE papers are "private communication", there are a few errors in the
number of resonances reported in Refs. \cite{Bo83}, and this reference does
not explain why some of the nuclides in their NDE were excluded from their
analysis of the neutron widths. For example, in Ref. \cite{Bo83} it is
stated that 1182 neutron widths in 21 nuclides were included in the
analysis. However, the sum of the number of resonances in these 21 nuclides
reported in this same reference is actually 1194. One resonance in $^{182}$W
was reported without a neutron width in the primary reference \cite{Ca73},
so only 40 of the 41 resonances for this nuclide can be included in the
width analysis. Therefore, the total number of widths is actually 1193, not
1182. Also, 19, 54, 47, and 21 resonances have been reported \cite{Ra74,Co79}
for $^{154,156,158,160}$Gd, respectively, not 19, 47, 21, and 54,
respectively as stated in Ref. \cite{Bo83}. In addition, there are three
nuclides ($^{160}$Dy, $^{164}$Dy, and $^{186}$W, with 18, 20, and 14
resonances, respectively) in the NDE of Ref. \cite{Bo83} that were not
included in their width analysis, even though neutron widths were available 
\cite{Li75,Ca73}.

In total then, the NDE of Ref. \cite{Bo83} should contain 1245 neutron
resonances in 24 nuclides for which widths have been reported. I obtained
resonance energies and neutron widths for these nuclides from the primary
references given in Table \ref{NDETable}, and cross checked these data with
information in the EXFOR/CSISRS \cite{EXFOR} database. Given the problems
noted above, I cannot be certain that I have analyzed the same data as those
in the NDE\ of Ref. \cite{Bo83}. However, using data from the primary
references should minimize any differences with Ref. \cite{Bo83} as well as
make it easier for others to reconstruct the data set I have used.

\begin{table}[tbp] \centering%
\caption{NDE nuclides.\label{NDETable}} 
\begin{tabular}{ccccccccc}
\hline\hline
Nuclide & Ref. & E$_{max}$ (keV) & \multicolumn{6}{c}{ML Results} \\ 
&  &  & \multicolumn{3}{c}{Minimum Threshold} & \multicolumn{3}{c}{\textit{p}%
-free Threshold} \\ 
&  &  & $T_{\max }$ & N$_{res}$ & $\nu _{min}$ & $T_{\max }$ & N$_{res}$ & $%
\nu _{pf}$ \\ \hline
$^{64}$Zn & \cite{Ga81,Ga81a} & 367.55 & 0.024 & 103 & 1.35 $%
_{-0.22}^{+0.24} $ & 0.05 & 99 & 1.54 $_{-0.26}^{+0.29}$ \\ 
$^{66}$Zn & \cite{Ga81,Ga81a} & 297.63 & 0.025 & 65 & 0.68 $_{-0.21}^{+0.23}$
& 0.05 & 61 & 0.74 $_{-0.25}^{+0.27}$ \\ 
$^{68}$Zn & \cite{Ga81,Ga81a} & 247.20 & 0.019 & 45 & 0.75 $_{-0.25}^{+0.27}$
& 0.05 & 41 & 0.95 $_{-0.32}^{+0.36}$ \\ 
$^{114}$Cd & \cite{Li74} & 3.3336 & 0.137 & 17 & 0.35 $_{-0.34}^{+0.54}$ & 
0.45 & 11 & 2.0 $_{-1.2}^{+1.5}$ \\ 
$^{152}$Sm & \cite{Ra72a} & 3.665 & 0.025 & 70 & 1.14 $_{-0.25}^{+0.27}$ & 
0.1 & 62 & 1.55 $_{-0.38}^{+0.40}$ \\ 
$^{154}$Sm & \cite{Ra72a} & 3.0468 & 0.036 & 27 & 0.76 $_{-0.36}^{+0.38}$ & 
0.1 & 22 & 1.32 $_{-0.55}^{+0.65}$ \\ 
$^{154}$Gd & \cite{Ra74} & 0.2692 & 0.15 & 19 & 0.44 $_{-0.43}^{+0.58}$ & 0.2
& 18 & 0.49 $_{-0.48}^{+0.64}$ \\ 
$^{156}$Gd & \cite{Co79} & 1.9908 & 0.009 & 54 & 1.22 $_{-0.26}^{+0.27}$ & 
0.2 & 46 & 1.44 $_{-0.49}^{+0.51}$ \\ 
$^{158}$Gd & \cite{Ra74} & 3.9827 & 0.012 & 47 & 0.75 $_{-0.22}^{+0.25}$ & 
0.2 & 35 & 1.17 $_{-0.47}^{+0.54}$ \\ 
$^{160}$Gd & \cite{Ra74} & 3.9316 & 0.013 & 21 & 0.55 $_{-0.33}^{+0.34}$ & 
0.2 & 16 & 0.83 $_{-0.65}^{+0.75}$ \\ 
$^{160}$Dy & \cite{Li75} & 0.4301 & 0.07 & 18 & 0.83 $_{-0.51}^{+0.57}$ & 0.2
& 14 & 1.41 $_{-0.83}^{+1.0}$ \\ 
$^{162}$Dy & \cite{Li75} & 2.9572 & 0.046 & 46 & 1.02 $_{-0.32}^{+0.33}$ & 
0.2 & 40 & 0.99 $_{-0.43}^{+0.47}$ \\ 
$^{164}$Dy & \cite{Li75} & 2.9687 & 0.04 & 20 & 0.82 $_{-0.44}^{+0.51}$ & 0.2
& 16 & 2.3 $_{-1.0}^{+1.2}$ \\ 
$^{166}$Er & \cite{Li72a} & 4.1693 & 0.02 & 109 & 0.85 $_{-0.17}^{+0.18}$ & 
0.3 & 78 & 1.85 $_{-0.45}^{+0.49}$ \\ 
$^{168}$Er & \cite{Li72a} & 4.6711 & 0.04 & 48 & 0.80 $_{-0.27}^{+0.30}$ & 
0.3 & 37 & 1.32 $_{-0.55}^{+0.62}$ \\ 
$^{170}$Er & \cite{Li72a} & 4.7151 & 0.05 & 31 & 0.36 $_{-0.31}^{+0.34}$ & 
.03 & 17 & 3.6 $_{-1.3}^{+1.6}$ \\ 
$^{172}$Yb & \cite{Li73} & 3.9000 & 0.05 & 55 & 0.71 $_{-0.26}^{+0.27}$ & 
0.06 & 54 & 0.70 $_{-0.26}^{+0.30}$ \\ 
$^{174}$Yb & \cite{Li73} & 3.2877 & 0.02 & 19 & 0.80 $_{-0.39}^{+0.44}$ & 
0.06 & 16 & 1.29 $_{-0.58}^{+0.68}$ \\ 
$^{176}$Yb & \cite{Li73} & 3.9723 & 0.015 & 23 & 0.04 $_{-0.03}^{+0.29}$ & 
0.06 & 15 & 1.05 $_{-0.55}^{+0.65}$ \\ 
$^{182}$W & \cite{Ca73} & 2.6071 & 0.069 & 40 & 0.76 $_{-0.35}^{+0.37}$ & 
0.15 & 34 & 1.50 $_{-0.55}^{+0.62}$ \\ 
$^{184}$W & \cite{Ca73} & 2.6208 & 0.04 & 30 & 0.62 $_{-0.31}^{+0.34}$ & 0.15
& 26 & 0.99 $_{-0.48}^{+0.54}$ \\ 
$^{186}$W & \cite{Ca73} & 1.1871 & 0.07 & 14 & 1.23 $_{-0.62}^{+0.78}$ & 0.15
& 13 & 1.32 $_{-0.75}^{+0.93}$ \\ 
$^{232}$Th & \cite{Ra72} & 2.988 & 0.016 & 178 & 0.76 $_{-0.12}^{+0.13}$ & 
0.26 & 123 & 1.78 $_{-0.34}^{+0.36}$ \\ 
$^{238}$U & \cite{Ra72} & 3.0151 & 0.0045 & 146 & 0.79 $\pm $ 0.12 & 0.47 & 
84 & 1.02 $_{-0.34}^{+0.39}$ \\ 
W.A. &  & - & - & 1245 & 0.801 $\pm $ 0.052 & - & 978 & 1.217 $\pm $ 0.092
\\ \hline\hline
\end{tabular}
\end{table}%

\section{Importance of threshold in ML analysis of neutron widths\label%
{Threshold}}

Because the PTD is a special case (degrees-of-freedom $\nu =1$) of the
family of $\chi ^{2}$ distributions, it is assumed that the data are
distributed accordingly and the ML method is used to estimate the most
likely value of $\nu $. Ideally, the data should be complete (no missing
resonances) and pure (all resonances have the same parity). Unfortunately,
all experiments from which neutron widths have been determined have finite
thresholds for detecting resonances and for separating small \textit{s}-
from large \textit{p}-wave resonances. Not properly accounting for these
effects can result in substantial systematic errors in ML estimates of $\nu $%
. Given the shape of the $\chi ^{2}$ distribution as a function of $\nu $,
neglecting the effect of missed \textit{s}-wave resonances below threshold
will, in general, lead to a falsely large value of $\nu $ from the ML\
analysis. Conversely, including even just a few \textit{p}-wave resonances
in an \textit{s}-wave set will, in general, lead to a falsely small value of 
$\nu $. This potential problem is especially important for many of the NDE
nuclides because they are near the peaks of the \textit{p}- and minimum of
the \textit{s}-wave neutron strength functions, and therefore neutron width
is a much less reliable indicator of resonance parity.

\begin{figure}[tbp]
\includegraphics*[width=100mm,keepaspectratio]{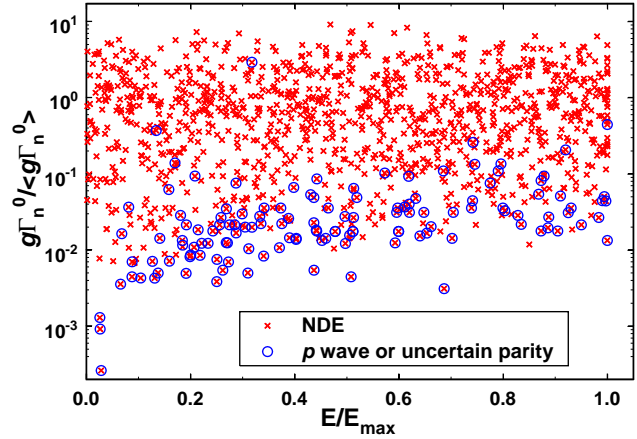}%
\caption{(Color online) All 1245 reduced neutron widths in the NDE (red
X's). Data for each nuclide have been normalized to their respective average
reduced neutron widths and maximum energies. Blue circles depict those
resonances which have been identified as being \textit{p} wave or of
uncertain parity.}
\label{NDEAllGn0VsEwPwave}
\end{figure}

\begin{figure}[tbp]
\includegraphics*[width=100mm,keepaspectratio]{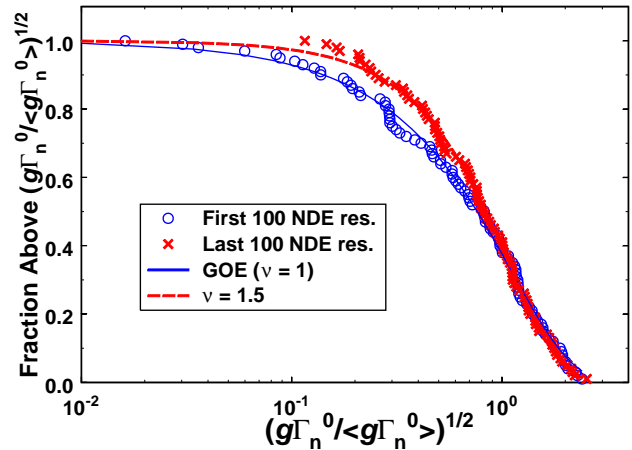}%
\caption{(Color online) Reduced neutron width distributions for the 100
lowest- (blue cirles) and highest- (red X's) energy resonances in the NDE.
The blue solid curve depicts the RMT prediction for the GOE (the PTD)
whereas the red dashed curve is for a $\protect\chi ^{2}$ distribution
having $\protect\nu $=1.5. See text for details.}
\label{NDEAllF100VsL100Chi2}
\end{figure}

Below I will show that the NDE is seriously contaminated by \textit{p}-wave
resonances. It is also almost certainly incomplete, as illustrated in Figs. %
\ref{NDEAllGn0VsEwPwave} and \ref{NDEAllF100VsL100Chi2}. Reduced neutron
widths for all 1245 resonances in the NDE are shown as a function of
resonance energy in Fig. \ref{NDEAllGn0VsEwPwave}, from which it can be seen
that their are fewer small widths as the energy increases. This is just what
is expected from well-known \cite{Fu65} instrumental effects that decrease
sensitivity as the energy increases; hence, more small resonances are missed
at higher energies. This is further illustrated in Fig. \ref%
{NDEAllF100VsL100Chi2} in which distributions of reduced neutron widths are
shown for the 100 lowest- and highest-energy NDE resonances. As can be seen
in this figure, the distribution of the highest-energy set is substantially
narrower (and hence is in better agreement with a $\chi ^{2}$ distribution
having a larger $\nu $) than the lowest energy one. Again, this is just what
is expected if more resonances are missed as energy increases.

Typically, these difficulties have been surmounted by using a
energy-independent threshold as an integral part of the ML analysis,
implicitly assuming that all \textit{s}-wave resonances above threshold have
been observed. We recently have shown \cite{Ko2010a} that an
energy-dependent threshold (on $g\Gamma _{n}^{0}$) of the form $T=a\,E_{%
\mathrm{n}}$, where $a$ is a~constant factor, offers three advantages
compared to using a energy-independent threshold. First, \textit{p}-wave
contamination is eliminated equally effectively at all energies. This is
because the penetrability factor for \textit{p}-waves differs from \textit{s}%
-waves by (to good approximation) a factor of $E_{\mathrm{n}}$. Second,
experiment thresholds have approximately this same energy dependence; thus,
possible diffusiveness of the instrumental threshold can be surmounted
equally effectively at all energies. Third, statistical precision of the
analysis is maximized by allowing the largest \textit{p}-wave-free set of 
\textit{s}-wave resonances to be included.

\begin{figure}[tbp]
\includegraphics*[width=100mm,keepaspectratio]{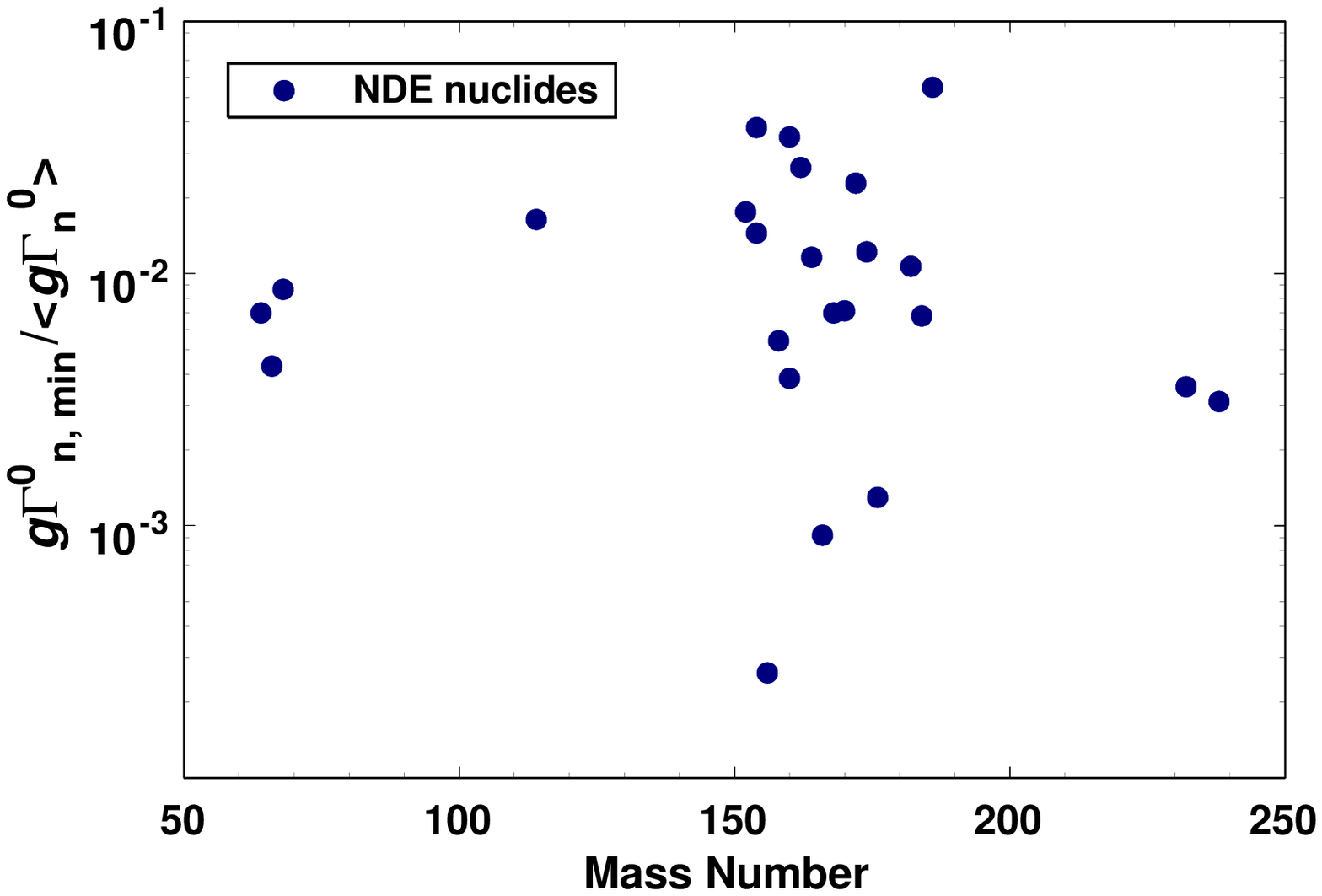}%
\caption{(Color online) Blue circles depict minimum reduced neutron widths,
normalized to their respective averages, for each of the nuclides in the
NDE, versus mass number.}
\label{MinWidthVsA}
\end{figure}

Analyzing a data set comprised of many different nuclides such as the NDE
involves at least two additional potential pitfalls. First, as is evident
from Fig. \ref{MinWidthVsA}, the apparent sensitivities of the various
experiments from which the NDE was derived differ by several orders of
magnitude. Therefore, if the entire NDE were analyzed as a single set (as
was done in Ref. \cite{Bo83}), the threshold must be at least as high as the
highest apparent individual threshold. But doing this will substantially
reduce the statistical precision of the result and at least partially negate
the reason for assembling the NDE in the first place. Second, for a given
set of data and threshold, the ML-estimated average reduced neutron width
might be substantially different from the one estimated from the data (e.g.,
from a simple average or by assuming $\nu =1$). Therefore, it is important
to include the average reduced width as a parameter in the ML\ analysis.
Furthermore, the difference between these two estimated values will likely
vary from one NDE nuclide to the next. For these two reasons, it is
important that separate ML analyses be made for each NDE nuclide.

To minimize the effects of the above problems, I have done separate ML
analyses for each nuclide in the NDE using (a range of) separate
energy-dependent thresholds, and then combined these individual results for
comparison to theory.

\section{An improved ML analysis of the NDE neutron widths\label{MLanalysis}}

The analysis technique was briefly described in Ref. \cite{Ko2010a}. Each
resonance $\lambda $ has an energy $E_{\lambda }$ and reduced neutron width $%
g\Gamma _{\lambda \mathrm{n}}^{0}$. The probability density function (PDF) $%
f(x|\nu )$ for a $\chi ^{2}$ distribution is given by,

\begin{equation}
f(x|\nu )dx=\frac{\nu }{2G(\frac{\nu }{2})}(\frac{\nu x}{2})^{\frac{\nu }{2}%
-1}\exp (-\frac{\nu x}{2})dx,  \label{Chi2DistEq}
\end{equation}%
where $G(\frac{\nu }{2})$ is the gamma function for $\frac{\nu }{2}$, and $%
x\rightarrow g\Gamma _{\lambda n}^{0}/\mathrm{E}[g\Gamma _{\lambda n}^{0}]$,
where $\mathrm{E}[g\Gamma _{\lambda n}^{0}]$ is the expectation (average)
value of the reduced neutron width, with $\mathrm{E}[\bullet ]$ denoting the
expectation value operator. The ML method is used to estimate most likely
values of $\nu $ and $\mathrm{E}[\Gamma _{\lambda \mathrm{n}}^{0}]$ as well
as their uncertainties.

To facilitate comparison of the various nuclides in the NDE (e.g., Figs.\ref%
{NDEAllGn0VsEwPwave}, \ref{NDEAllF100VsL100Chi2}, \ref{MinWidthVsA}, and \ref%
{CdErThGn0vEAndNuvT}, and Table \ref{NDETable}), I have normalized data for
each nuclide to their respective average reduced widths $<g\Gamma _{n}^{0}>$
(as reported in the primary references or in Ref. \cite{Mu2006}) and maximum
energies $E_{\max }$. Hence, thresholds (on $g\Gamma _{n}^{0}/<g\Gamma
_{n}^{0}>$) can be expressed as $T^{\prime }=T_{\max }\,E_{\mathrm{n}%
}/E_{\max }$ and thus, $T_{\max }$ is the maximum value of the threshold
relative to the average reduced neutron width. Values of $T_{\max } $ used
in the analyses are given in Table \ref{NDETable}.

The joint PDF for statistical variables $g\Gamma _{\lambda \mathrm{n}}^{0}$
and $E_{\lambda }$ is defined in a~2D region ${\mathcal{I}}$ given by
inequalities $E_{\lambda }<E_{\mathrm{max}}$ and $g\Gamma _{\lambda \mathrm{n%
}}^{0}>T(E_{\lambda })$. The expression for this PDF reads%
\begin{equation}
h^{0}\!\left( E_{\lambda },g\Gamma _{\lambda \mathrm{n}}^{0}\,|\,\nu ,%
\mathrm{E}[g\Gamma _{\lambda \mathrm{n}}^{0}]\right) =Cf\!\left( \left. 
\frac{g\Gamma _{\lambda \mathrm{n}}^{0}}{\mathrm{E}[g\Gamma _{\lambda 
\mathrm{n}}^{0}]}\right\vert \nu \right) .
\end{equation}%
The factor $C$, ensuring a~unit norm of $h^{0}$, is $\nu $- and $\mathrm{E}%
[g\Gamma _{\lambda \mathrm{n}}^{0}]$-dependent. The ML function was
calculated from all $n_{0}$ pairs $\left[ E_{\lambda _{i}}^{\mathrm{\,exp}%
},g\Gamma _{\lambda _{i}\mathrm{n}}^{\mathrm{\,exp}}\right] $, obtained from
the experiment, which fall into the region $\mathcal{I}$. Specifically, 
\begin{equation}
L\left( \nu ,\mathrm{E}[g\Gamma _{\lambda \mathrm{n}}^{0}]\right)
=\prod_{i=1}^{n_{0}}h^{0}\!\left( E_{\lambda _{i}}^{\mathrm{\,exp}},g\Gamma
_{\lambda _{i}\mathrm{n}}^{\mathrm{0\,exp}}\,|\,\nu ,\mathrm{E}[g\Gamma
_{\lambda \mathrm{n}}^{0}]\right) .
\end{equation}

For the initial analyses, thresholds just below the smallest observed
resonance for each nuclide were used. ML results with these thresholds are
given in column $\nu _{\min }$ of Table \ref{NDETable}. Because experiment
thresholds might not be precisely sharp, it is expected that the resulting $%
\nu _{\min }$ values would be systematically a bit large. However, almost
all $\nu _{\min }$ values are less than the PTD value of 1.0 .

\begin{figure}[tbp]
\includegraphics*[width=160mm,keepaspectratio]{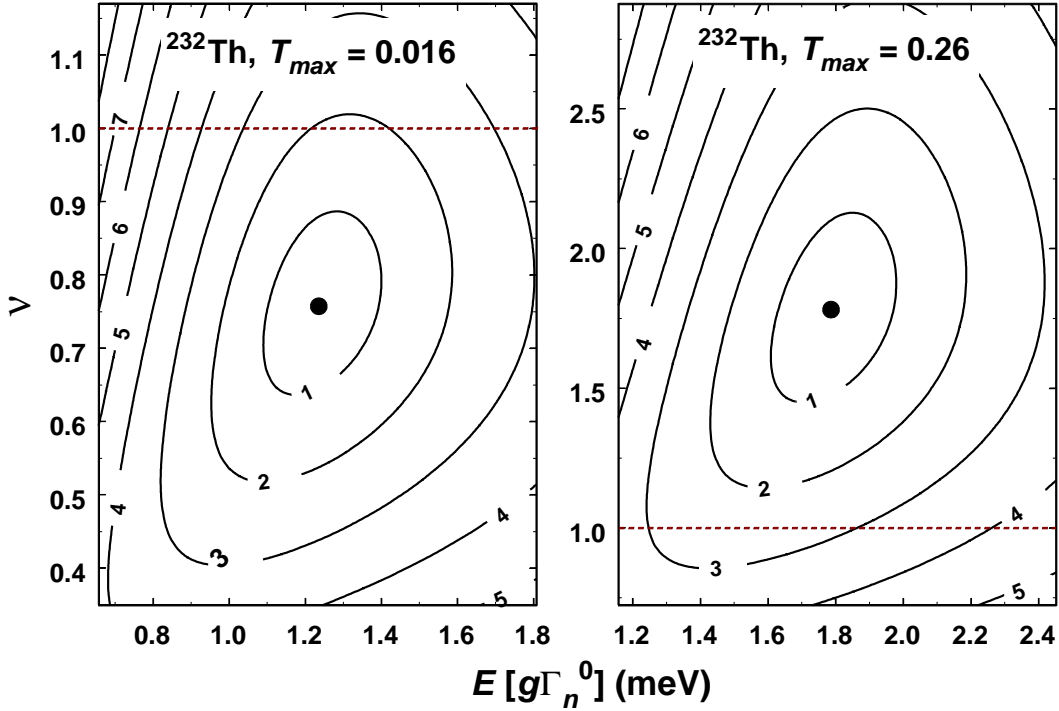}%
\caption{Plots of $z(\protect\nu ,E[g\Gamma _{\protect\lambda n}^{0}])$
constructed from ML analyses, at minimum (left) and \textit{p}-wave-free
(right) thresholds, of the NDE $^{232}$Th data. On each plot, a filled
circle indicates the location of $L_{\max }$, and a dashed horizontal line
is drawn at $\protect\nu =1$, the PTD value.}
\label{Th232ML2DConT0p016}
\end{figure}

Contour plots of ML functions for $^{232}$Th, calculated at two different
thresholds, in the form%
\begin{equation}
z\!\left( \nu ,\mathrm{E}[g\Gamma _{\lambda \mathrm{n}}^{0}]\right) =2^{%
\frac{1}{2}}\left[ \ln L_{\mathrm{max}}-\ln L\left( \nu ,\mathrm{E}[g\Gamma
_{\lambda \mathrm{n}}^{0}]\right) \right] ^{\frac{1}{2}}
\end{equation}%
are shown in Fig.~\ref{Th232ML2DConT0p016}. Here, $L_{\mathrm{max}}$ is the
maximum of the ML function. Contours at fixed $z=k$ encircle approximately
the $k\sigma $ confidence region, and were used to derive the $1\sigma $
uncertainties given in Table \ref{NDETable}. Careful statistical analysis in
Ref. \cite{Ko2010a} verified that these contours are reliable. The weighted
average of results at minimum threshold for the 24 nuclides in the NDE is $%
\nu =0.801\pm 0.052$, which is 3.8 standard deviations smaller than the
predicted result of $\nu =1$. Hence, these data reject the PTD with a
statistical significance of 99.98\%. To check this result, the combined
probability for the 24 NDE nuclides was calculated using Fisher's and
Stouffer's (both weighted and unweighted) techniques \cite{Wh2005}. These
methods yielded combined confidence levels of 99.97\%, 99.98\%, and 99.99\%,
respectively, in good agreement with the weighted-average result. Hence, at
minimum thresholds the NDE data reject the PTD with high confidence and
indicate that $\nu $ is significantly less than 1.0. A $\nu $ value
significantly less than the PTD could be a sign of interesting physics \cite%
{Al92,Ko2010a,Al89,Ro2010,Ce2011,Vo2011}. However, a more likely explanation
is that the NDE contains sizeable \textit{p}-wave contamination.

\section{Cleansing the NDE of \textit{p}-waves\label{Ridding}}

\begin{figure}[tbp]
\includegraphics*[width=100mm,keepaspectratio]{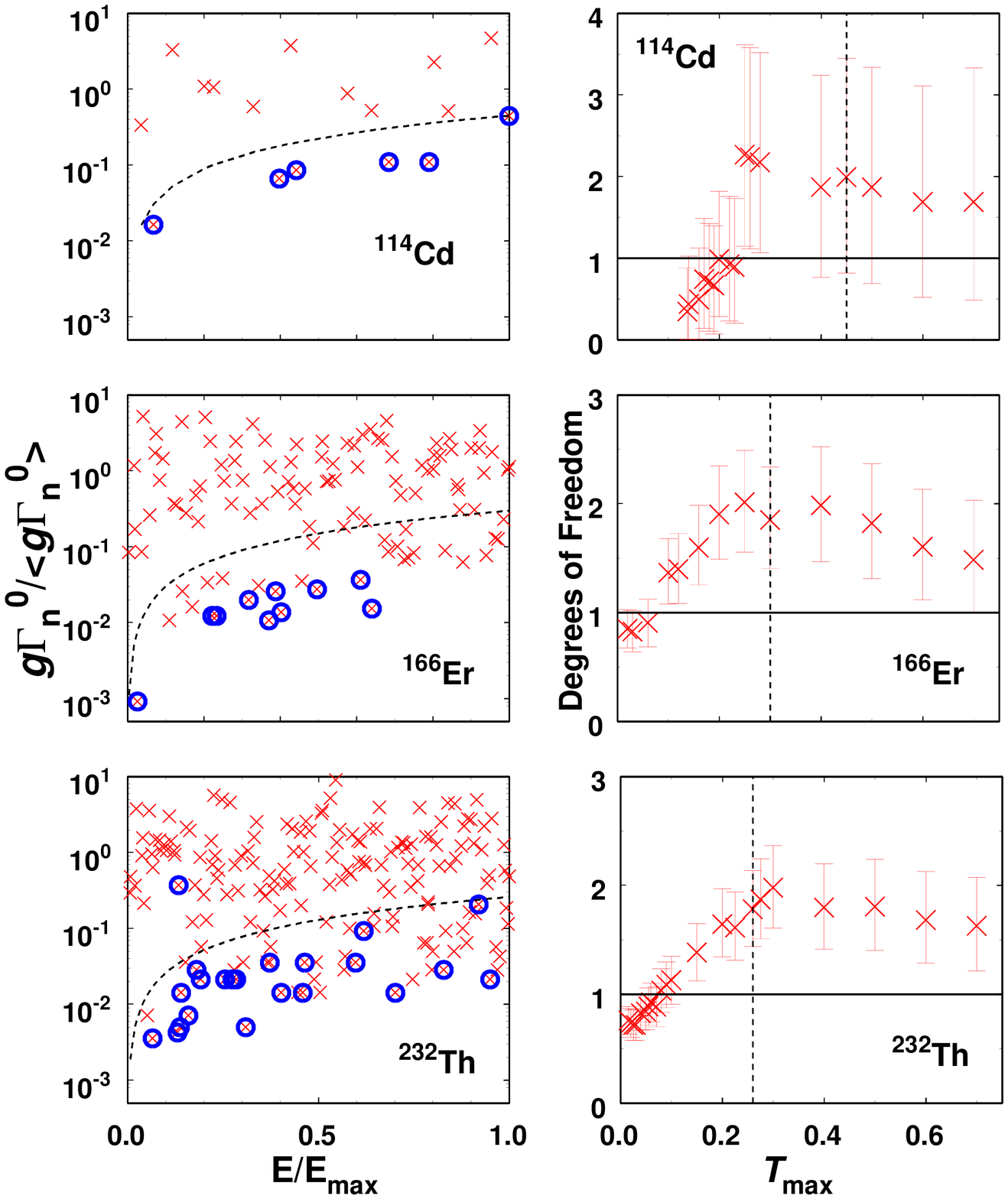}%
\caption{(Color online) Left: normalized reduced neutron widths versus
normalized resonance energies for $^{114}$Cd, $^{166}$Er, and $^{232}$Th
resonances in the NDE. Red X's depict all resonances in the NDE whereas blue
circles show resonances previously identified as being \textit{p} wave or of
uncertain parity. Right: Red X's depict $\protect\nu $ values from ML
analyses versus thresholds used, for the same three nuclides. Error bars
represent $1\protect\sigma $ confidence levels. Black dashed vertical lines
correspond to thresholds depicted by black dashed curves in the left part of
this figure. See text for details.}
\label{CdErThGn0vEAndNuvT}
\end{figure}

Results of the maximum-likelihood analyses for three NDE nuclides are shown
in Fig. \ref{CdErThGn0vEAndNuvT}. On the left of this figure, normalized
reduced neutron widths are plotted as functions of normalized resonance
energy. On the right side of this figure, $\nu $ values from the ML analyses
are plotted versus threshold coefficients $T_{\max }$.

\begin{figure}[tbp]
\includegraphics*[width=100mm,keepaspectratio]{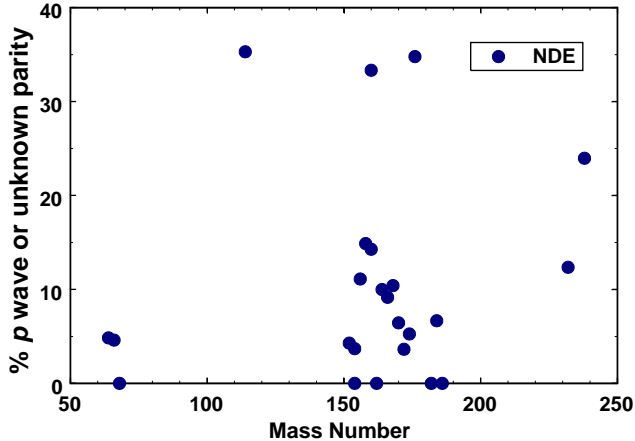}%
\caption{(Color online) Blue circles depict percentages of NDE resonances
for each nuclide which are known to be \textit{p} wave or are of uncertain
parity, as a function of mass number.}
\label{NDEAllPerPvsA}
\end{figure}

That the NDE is contaminated by \textit{p}-wave resonances is evident from
this figure in two ways and from Figs. \ref{NDEAllGn0VsEwPwave} and \ref%
{NDEAllPerPvsA}. First, resonances in the NDE that have been identified (in
Refs. \cite{Su98,Mu2006} and references contained therein) as \textit{p}%
-wave or of uncertain parity are shown as open circles in Figs. \ref%
{NDEAllGn0VsEwPwave} and \ref{CdErThGn0vEAndNuvT}. For example, in Ref. \cite%
{Co78}, 58 \textit{p}-wave resonances in $^{232}$Th were assigned on the
basis of $\gamma $-cascade information, 13 of which are in the NDE. One of
these 13 resonances also is known \cite{Fr92,St98} to be \textit{p} wave by
its observed parity-violating asymmetry. Percentages of \textit{p}-wave or
unknown-parity resonances for NDE nuclides, as a function of mass number,
are shown in Fig. \ref{NDEAllPerPvsA}. For over half (14/24) of the NDE
nuclides analyzed herein, these resonances account for 5\% or more of the
total, for 10 of the 24 they are at least 10\%, and in the three worst cases
about 35\%.

Second, that many of the NDE resonances are in fact \textit{p}-wave is
reinforced by the behavior of the $\nu $ values from the ML analyses as
functions of threshold, as shown in the right side of Fig. \ref%
{CdErThGn0vEAndNuvT}. In all three cases shown, $\nu $ systematically
increases with threshold before gradually stabilizing. This is just the
behavior expected for a population of \textit{s}-wave resonances
contaminated by \textit{p}-wave resonances. Similar fractions of previously
identified \textit{p}-wave resonances and trends in $\nu $ with threshold
are seen for several of the other NDE nuclides.

Removing effects of these \textit{p}-wave resonances from the NDE ML
analysis is a simple matter of raising thresholds until they are above the
largest previously-identified \textit{p}-wave resonance and/or $\nu $
stabilizes as a function of threshold. Typical "\textit{p}-wave free"
thresholds for three NDE nuclides are shown as dashed curves in the
left-hand part of Fig. \ref{CdErThGn0vEAndNuvT}, and corresponding values of 
$T_{\max }$ are depicted by dashed vertical lines in the right-hand part of
this figure. Degrees-of-freedom values for each of the NDE nuclides at these
"\textit{p}-wave free" thresholds are given in column five ($\nu _{pf}$) of
Table \ref{NDETable}. The resulting weighted average for the NDE is still in
conflict with the RMT prediction for the GOE, albeit in the opposite
direction from the result using the lowest thresholds: $\nu =1.217\pm 0.092$%
, corresponding to a confidence level of 98.17\% for excluding the PTD.
Fisher's and unweigthed and weighted Stouffer's confidence levels are
somewhat higher; 99.15\%, 99.81\%, and 99.37\%, respectively. Hence, when
the NDE is cleansed of \textit{p}-wave resonances, the data still reject the
PTD with high confidence.

\section{Discussion\label{Why}}

I have shown that when neutron widths in the NDE are analyzed carefully and
in such a way as to eliminate \textit{p}-wave resonances, the data exclude
the PTD with fairly high confidence ($\nu =1.217\pm 0.092$). At the same
time, it has been shown \cite{Ha82,Bo83} that the NDE as a whole agrees
remarkably well with energy-level fluctuations predicted by RMT for the GOE.
Given the greater sensitivity of widths over energies to the degree of chaos
in the system \cite{Al92}, and to effects related to the "openess" of the
system \cite{Kl85,Ro2010,We2010,Ce2011,Vo2011}, perhaps this dichotomy can
be resolved. However, I am not aware of any published model which does so
while at the same time yielding $\nu >1$.

For example, in the calculations of Ref. \cite{Al92}, collective effects
sometimes resulted in regions of model space where transition-strength
distributions deviated strongly from the PTD but, at the same time,
energy-level fluctuations were consistent with GOE predictions. However,
transition-strength distributions in this model deviated from the PTD in the
direction opposite ($\nu <1$) of the NDE.

Reaction effects also should be considered to explain the dichotomy. For
example, it recently was proposed \cite{We2010} that the standard
transformation of measured to reduced \textit{s}-wave neutron widths, $%
\Gamma _{n}^{0}=\Gamma _{n}/\sqrt{E}$, should be modified for nuclides near
the peaks of the \textit{s}-wave neutron strength function. This newly
proposed transformation, $\Gamma _{n}^{0}=\Gamma _{n}/\sqrt{E}\times \pi (%
\frac{\text{%
h{\hskip-.2em}\llap{\protect\rule[1.1ex]{.325em}{.1ex}}{\hskip.2em}%
}^{2}}{2m})^{1/2}(E+|E_{0}|)$, where $E_{0}$ is the energy (relative to
threshold) of the \textit{s}-wave single-particle state responsible for the
peak in the neutron strength function, could affect (relative to the
standard transformation) the shape of the $\Gamma _{n}^{0}$ distribution for
values of $E_{0}$ near the resolved-resonance energy range. However, as most
of the NDE nuclides are not near the peaks of the \textit{s}-wave neutron
strength function, this proposal should not affect the NDE, and therefore
cannot explain the dichotomy.

Several explanations related to the "openess" of the system \cite%
{Kl85,Ro2010,Ce2011} and to correlations between incoming and outgoing
channels \cite{Vo2011} have been proposed for the observation \cite{Ko2010a}
that $\Gamma _{n}^{0}$ distributions for $^{192,194}$Pt resonances are
significantly broader ($\nu \approx 0.5$) than the PTD. However, width
distributions \cite{Ce2011,Vo2011} predicted by these theories are better
characterized by $\nu <1$ rather than $\nu >1$ as observed for the NDE.

Considering how data in the NDE were selected suggests another solution. It
often has been stated that $\Delta _{3}$ is very sensitive to missing or
misassigned resonances (see e.g., \cite{Mu2011,Mu2009,Mu2007} for recent
work on this subject). Therefore, if the NDE was pure but incomplete, or
vice versa, it would not be expected to agree well with the spacing
statistics used in Refs. \cite{Ha82,Bo83}. However, it seems plausible that
because the NDE is both impure and incomplete, it can be made to agree with
these statistics, especially considering that there are expected to be many
more \textit{p}- than \textit{s}-wave resonances at the small widths where
it was not possible (using means independent of RMT) to differentiate the
two parities, and hence an extremely large number (see below) of possible "%
\textit{s}-wave" sets can be constructed from the observed resonances.

Many different selections were applied to obtain the NDE. For example, in
Ref. \cite{Bo83} it is stated that "The criterion for inclusion in the NDE
is that the individual sequences be in general agreement with the GOE."
Furthermore, data from many of the included nuclides were selected, at least
in part, using measures derived from RMT for the GOE.

Data for all but three of the 24 nuclides considered herein were obtained by
the group at Columbia University. According to their publications (e.g., Ref.%
\cite{Li72a}), they had "...no specific tests for \textit{s} vs \textit{p}
levels, so there may be errors in these assignments." Therefore, they relied
on theoretical guidance, specifically measures derived from the GOE, to
perform these separations. For example, for six of the 24 nuclides
considered herein, including the two having the largest number of
resonances, separation of \textit{p}- from \textit{s}-wave resonances was
accomplished \cite{Li72} by i) assuming all resonances having neutron widths
larger than a certain (unspecified) size were \textit{s} wave, ii)
calculating the number of \textit{s}-wave resonances below this size by
assuming the PTD is correct, and iii) deciding which of the resonances below
the threshold defined in the first step and needed to achieve the total
number calculated in the second step, were \textit{s}-wave by requiring good
agreement with four spacing statistics (the Wigner nearest-neighboor spacing
distribution, $\rho (S_{j},S_{j+1})$, $\Delta _{3}$, and the Dyson $F$ test)
derived from the GOE. Separation of \textit{s}- from \textit{p}-wave
resonances for several of the other NDE nuclides followed the first two
steps described above followed by a Bayesian analysis to decide which of the
resonances below the threshold defined in the first step to assign to the 
\textit{s}-wave set. The Bayesian analysis again assumes that the PTD is
correct (for both \textit{s-} and \textit{p}-wave resonances) and
furthermore requires the average widths for \textit{s}- and \textit{p}-wave
resonances. The latter quantity usually is known only approximately, if at
all. Such Bayesian analyses are known to be unreliable. For example, several
neutron resonances in $^{64}$Zn \cite{Ga81} are known to be definitely 
\textit{p} wave by their symmetrical shape in transmission (total cross
section) data, but nevertheless have a Bayesian probability of $>$99 \% of
being \textit{s} wave.

Given these selection procedures then, it is perhaps understandable that the
NDE appears to agree well with the spacing statistics examined in Refs. \cite%
{Ha82,Bo83} despite that fact that the data are neither complete nor pure.
Consider, for example, the case of $^{232}$Th, the NDE nuclide with the
largest number of resonances. The authors of the primary reference from
which these data were taken \cite{Ra72,Li72} state that 80\% of the \textit{s%
}-wave set (i.e. the set included in the NDE) were chosen to be \textit{s}
wave because they were too large to be \textit{p} wave. Hence, 36 of the 178 
$^{232}$Th resonances fall below this size and had to be selected as \textit{%
s} wave using measures derived from the GOE, as noted above. In this same
reference, 62 resonances are assigned as \textit{p} wave (and hence by
definition are below the \textit{s}-wave threshold). Therefore, there are 98
resonances from which to choose the 36 needed to complete the \textit{s}%
-wave set. Therefore, the number of possible \textit{s}-wave sets is
astronomically large (\symbol{126}10$^{27}$), and hence it should not be
surprising that at least one set can be found to agree with the various RMT
measures used in Ref. \cite{Li72}. Similar numbers apply to the other data
sets from Columbia.

Finally, given that a significant fraction of the NDE data were selected
using measures derived from RMT for the GOE, it seems highly questionable to
use the full NDE for any test of this same theory. In contrast, the "\textit{%
p}-free" results presented herein circumvent this problem by eliminating
these problematical resonances from the test.

\section{Conclusions}

I have shown that neutron widths in the NDE \cite{Bo83}, when analyzed
carefully, reject the PTD with high confidence. Given that reported
deviations from the PTD for individual nuclides \cite%
{Ri69,Fo71,Fo71a,Ra72,Ca76a,Ko2007,Ko2010a} have occurred on both sides of $%
\nu =1$, the statistical tests used herein likely underestimate the
confidence with which the PTD can be rejected for the combined set of
nuclides in the NDE. Furthermore, I have shown that the NDE is not pure and
very likely incomplete. Also, measures derived from RMT for the GOE often
were used in deciding which resonances to include in the NDE. These facts
cast very serious doubt on repeated claims that the NDE constitutes a
striking confirmation of RMT.

Measurement techniques have improved considerably since the data in the NDE
were acquired. In particular, new methods for determining resonance spins 
\cite{Ko2007,Ko2009} hold the promise of surmounting the difficulty of
separating small \textit{s}- from large \textit{p}-wave resonances, which
has been one of the most troublesome barriers to obtaining better data.
Unfortunately most experimentalists continue to use RMT to correct their
data for instrumental effects rather than use their data to test RMT.
Exceptions to this practice are recent tests \cite{Ko2007,Ko2010a} which
have revealed significant disagreements between new neutron data and the
PTD. These new data, together with previously reported disagreements \cite%
{Ri69,Fo71,Fo71a,Ra72,Ca76a}, which have largely been ignored, are
potentially very interesting. Therefore, I urge experimentalists obtaining
new and improved data to use them, when possible, to test RMT.

\begin{acknowledgments}
This work was supported by the Office of Nuclear Physics of the U.S.
Department of Energy under Contract No. DE-AC05-00OR22725 with UT-Battelle,
LLC.
\end{acknowledgments}

\bibliographystyle{prsty}
\bibliography{ACOMPAT,pauls}

\newif\ifabfull\abfulltrue

\end{document}